\documentstyle[sprocl,psfig]{article}

\newcommand{\mafigura}[4]{
\begin{figure}[t]
\begin{center}
\psfig{figure=#2,width=#1}
\end{center}
\caption{#3 \label{#4}}
\end{figure} }
\newcommand{\eq}[1]{Eq.~(\ref{#1})}

\newcommand{\beq}{\begin{equation}}
\newcommand{\eeq}{\end{equation}}
\newcommand{\bea}{\begin{eqnarray}}
\newcommand{\eea}{\end{eqnarray}}

\newcommand{\yc}{y_{c}}

\newcommand{\ql}{\ell}

\begin{document}

\title{IMPROVED DETERMINATION OF THE $b$-QUARK MASS AT THE $Z$ PEAK
\footnote{Talk given at the IVth International
Symposium on Radiative Corrections (RADCOR98),
Barcelona, Catalonia, Spain, 8-12 Sep 1998.}}

\author{G. RODRIGO}

\address{INFN-Sezione di Firenze, Largo E. Fermi 2, 50125 Firenze
Italy\\E-mail: rodrigo@fi.infn.it} 

\author{A. SANTAMARIA}

\address{Department de F\'{\i}sica Te\`orica, IFIC, CSIC-Universitat 
de Val\`encia, 46100 Burjassot, Val\`encia, Spain}

\author{M. BILENKY}

\address{Laboratory of Nuclear Problems, JINR, 141980 Dubna,
Russian Federation}


\maketitle\abstracts{Next-to-leading order calculations of heavy quark
three-jet production in $e^+ e^-$ annihilation are reviewed.
Their application for the measurement of the $b$-quark mass
at LEP/SLC and to the test of the flavour independence 
of the strong coupling constant are discussed.
Prospects for future improvements are studied.}


\section{Introduction}

Effects of the bottom-quark mass,
$m_b$, were already noticed in the early tests~\cite{lep} 
of the flavour independence of the strong coupling constant,
$\alpha_s$, in $e^+e^-$-annihilation at the $Z$-peak. 
Motivated by the remarkable sensitivity of the three-jet observables
to the value of the quark mass, the possibility of the determination
of $m_b$ at LEP, assuming universality of the strong interactions,
was considered~\cite{Bilenkii:1995ad}.
This question was analyzed in detail in~\cite{Bilenkii:1995ad}, where
the necessity of the next-to-leading order (NLO) calculation
for the measurements of $m_b$ was also emphasized.

The NLO calculation for the process $e^+e^- \rightarrow 3jets$, with
complete quark mass effects, has been
performed independently by three groups~\cite{rsb97,bbu97,no97}.
These predictions are in agreement with each other
and were successfully used in the measurements of the $b$-quark
mass far above threshold~\cite{delphi97,sld}
and in the precision tests of the universality of the strong
interaction~\cite{delphi97,opal,sld} at the $Z$-pole.
In this talk we make a short review of such calculations.
Furthermore, prospects for future improvements are discussed.

It is surprising that at high energies the bottom-quark mass
could be relevant, since it appears screened by the center of
mass energy, $m_b^2/m_Z^2 \simeq 10^{-3}$ at LEP.
Nevertheless, when more exclusive processes than a total
cross section are considered, like a n-jet cross section,
mass effects can be enhanced as $m_b^2/m_Z^2/y_c$, where
$y_c$ is the parameter that defines the jet multiplicity.

Since quarks are {\it not free} particles
and, therefore, their mass can be considered like a coupling
constant, one has the freedom to use different quark mass definitions,
e.g. the perturbative pole mass $M_b$ or the $\overline{MS}$
scheme running mass $m_b(\mu)$.
Physics should be independent of it but at a fixed order in
perturbation theory there is a significant dependence on which
mass definition is used, as well as on the renormalization scale $\mu$.
The inclusion of higher orders to reduce these two
uncertainties, due to mass definition and $\mu$ scale, is mandatory
for an accurate description of mass effects.


\section{Three jet observables and the measurement of $m_b$}

The observable proposed some time ago to measure
the bottom-quark mass at the $Z$-resonance was the 
ratio~\cite{Bilenkii:1995ad}
\beq
R^{bd}_3 \equiv \frac{\Gamma^b_{3j}(\yc)/\Gamma^b}
{\Gamma^d_{3j}(\yc)/\Gamma^d}~,
\label{eq:r3bd_def}
\eeq
where $\Gamma^q_{3j}$ and $\Gamma^q$ are the three-jet and the
total decay widths of the $Z$-boson into a quark pair of flavour $q$
in a given jet-clustering algorithm.
More precisely, the measured quantity is
\beq
R^{b\ql}_3 \equiv \frac{\Gamma^b_{3j}(\yc)/\Gamma^b}
{\Gamma^{\ql}_{3j}(\yc)/\Gamma^{\ql}} 
= 1 + \frac{\alpha_s(\mu)}{\pi} \: a_0(y_c) + 
r_b \left(b_0(r_b,y_c) + \frac{\alpha_s(\mu)}{\pi} \: b_1(r_b,y_c)\right)~,
\label{eq:pole}
\eeq
where now the sum of the contributions of the three light flavours
$\ql=u,d,s$ is included in the denominator.
The $R_3^{bd}$ and the $R_3^{b\ql}$ observables differ only by
the function $a_0$ (which is zero for $R_3^{bd}$).
This contribution originates from the triangle
diagrams~\cite{Hagiwara:1991dx}. It is numerically very small
(0.002 for the Durham jet algorithm) and
almost independent of the $b$-quark mass.
The $b_0$ and $b_1$ functions give respectively the leading-order (LO)
and NLO mass corrections, once the leading dependence
on $r_b = M_b^2/m_Z^2$, where $M_b$ is the bottom-quark pole mass,
has been factorized out.

\mafigura{10.5 cm}{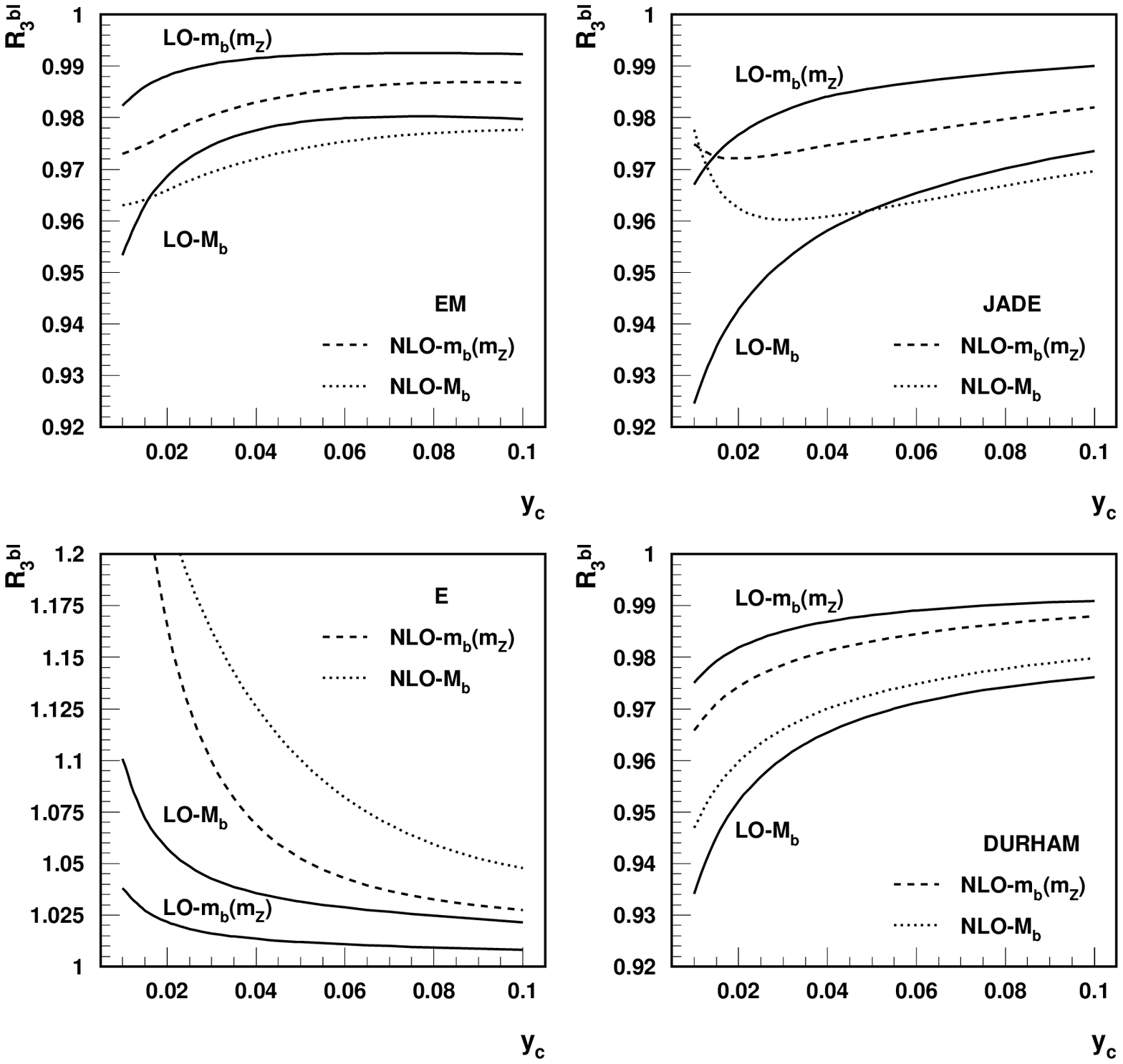}
{The observable $R_3^{b\ql}$ as a function of $\yc$ at the NLO for
the four algorithms considered. The dotted lines give the observable
computed at the NLO in terms of the pole mass $M_b=4.6$~GeV while
the dashed lines are obtained when it is written in terms of the
running mass $m_b(m_Z)=2.83$~GeV. In both cases the renormalization scale
is fixed at $\mu=m_Z$, and $\alpha_s(m_Z)=0.118$. For comparison we also
plot in solid lines the LO results for $M_b=4.6$~GeV (LO-$M_b$) and
$m_b(m_Z)=2.83$~GeV (LO-$m_b(m_Z)$).
}{fig:r3bl}

Ratios of differential two-jet rates, where 
the two-jet width $\Gamma_{2j}$ is calculated from
the three- and the four-jet fractions through the identity
$\Gamma_{2j} = \Gamma - \Gamma_{3j} - \Gamma_{4j}$, have
been studied in~\cite{Rodrigo:1998nk,bbu97}.
Ratios of event shape distributions have also been considered~\cite{no97}.

Using the known relationship between the pole mass and the 
$\overline{MS}$ scheme running mass,
\beq
M_b^2 = m_b^2(\mu) \left[1+\frac{2\alpha_s(\mu)}{\pi} 
\left(\frac{4}{3} -\log \frac{m_b^2}{\mu^2} \right)\right]~,
\label{eq:poltorunning}
\eeq
we can re-express~\eq{eq:pole} in terms of the running mass
$m_b(\mu)$. Then, keeping only terms of order ${\cal O}(\alpha_s)$
we obtain
\bea
R_3^{b\ql} &=& 1 + \frac{\alpha_s(\mu)}{\pi} \: a_0(y_c) + 
\bar{r}_b(\mu) \left( b_0(\bar{r}_b,y_c) + 
\frac{\alpha_s(\mu)}{\pi} \: \bar{b}_1(\bar{r}_b,y_c,\mu) \right)~, 
\label{eq:MS}
\eea
where $\bar{r}_b(\mu)=m_b^2(\mu)/m_Z^2$ and
$\bar{b}_1=b_1+ 2 b_0 (4/3 - \log r_b + \log(\mu^2/m_Z^2))$.
Although at the perturbative level both expressions, \eq{eq:pole}
and \eq{eq:MS}, are equivalent they give different answers since 
different higher order contributions have been neglected.
The spread of the results gives an estimate of the size of 
higher order corrections.

In fig.~\ref{fig:r3bl}  we present our results
for $R_3^{b\ql}$ in the four clustering algorithms:
EM~\footnote{A modification of the standard Jade scheme, convenient for
massive parton calculations~\cite{Bilenkii:1995ad}.}, Jade, E and Durham.
For all the algorithms we plot the NLO results written either
in terms of~\cite{pich} the pole mass, $M_b=4.6$~GeV, or
in terms of the running mass at $m_Z$, $m_b(m_Z)=2.83$~GeV.
The renormalization scale is fixed to $\mu=m_Z$ and
$\alpha_s(m_Z)=0.118$. For comparison we also show $R_3^{b\ql}$
at LO when the value of the pole mass, $M_b$, or
the running mass at $m_Z$, $m_b(m_Z)$, is used for the quark mass.

Note the different behaviour of the different algorithms.
In particular the $E$ algorithm.
As already discussed in~\cite{Bilenkii:1995ad},
in this algorithm the shift in the resolution parameter produced
by the quark mass makes the mass corrections positive while
by kinematical arguments one would expect a negative effect,
since massive quarks radiate less gluons than massless quarks.
Furthermore, the NLO corrections are very large in the $E$ algorithm
and strongly dependent on $y_c$.
All this probably indicates that 
it is difficult to give an accurate QCD prediction for it.
For the Jade algorithm the NLO correction written in terms of
the pole mass starts to be large for $\yc \le 0.02$.
Note, however that the NLO correction written in terms of the running
mass is still kept in a reasonable range in this region. 
Durham, in contrast, is the algorithm that presents a better
behaviour for relatively low values of $\yc$ while keeping
NLO corrections in a reasonable range. 

\mafigura{12 cm}{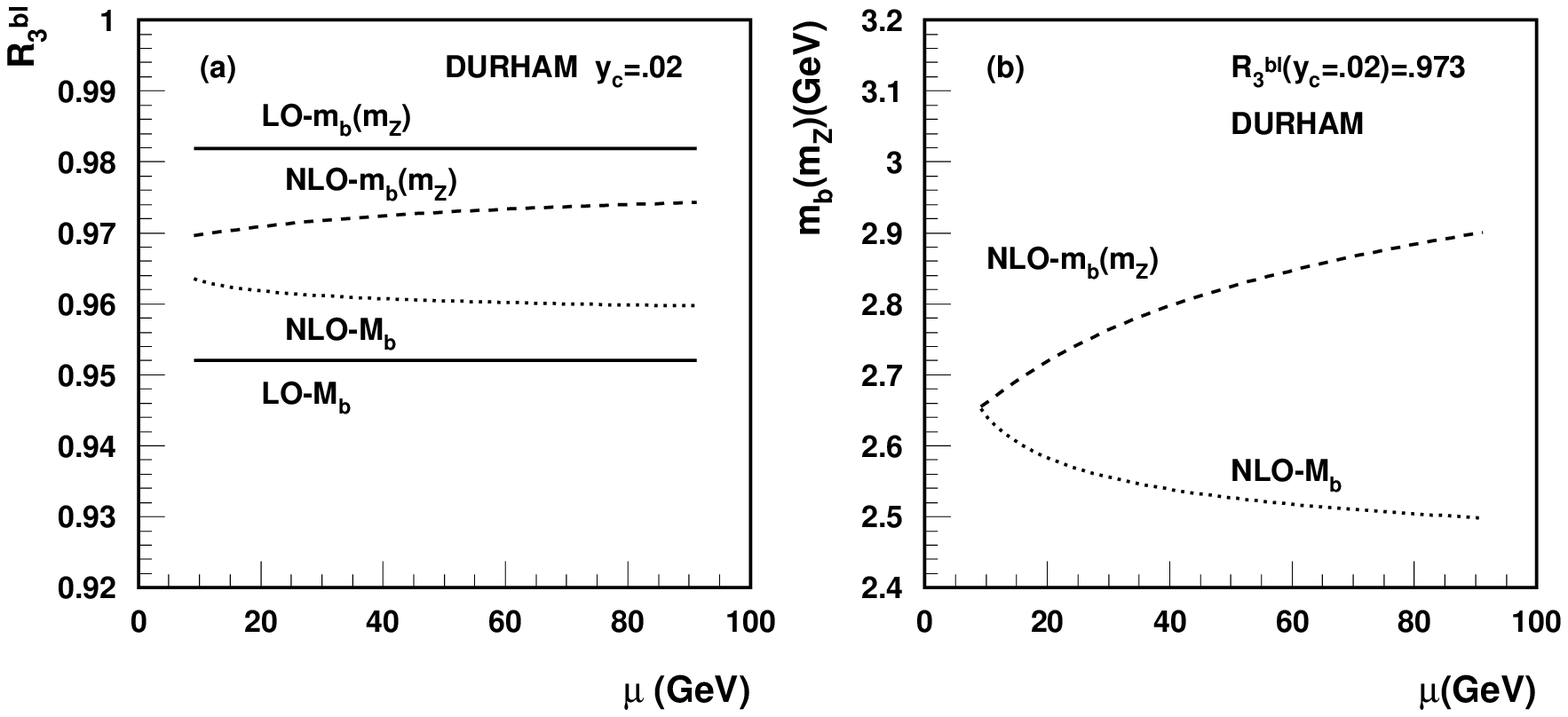}
{Durham algorithm:
(a) Renormalization scale dependence of the NLO
predictions written in terms of either the running mass, NLO-$m_b(m_Z)$,
or the pole mass, NLO-$M_b$, for the $R_3^{b\ql}$ ratio at a fixed
value of $y_c$.
(b) Predicted value of $m_b(m_Z)$ for a fixed value of $R_3^{b\ql}$
using either the running expression, NLO-$m_b(m_Z)$, or the
pole expression, NLO-$M_b$, as a function of the scale $\mu$.}
{fig:radmu_dur}

The theoretical predictions for the observables studied
contain a residual dependence on the
renormalization scale $\mu$. To give an idea of the uncertainties
introduced by this we plot in fig.~\ref{fig:radmu_dur}a
the observable $R_3^{b\ql}$ as a function of $\mu$ for a fixed value of 
$\yc$. Here we only present plots for the Durham algorithm, the one 
with the better behaviour. 
We use the following one-loop evolution equations 
\beq 
a(\mu) = \frac{a(m_Z)}{K}~,
\qquad
m_b(\mu) = m_b(m_Z) \: K^{-\gamma_0/\beta_0}~,
\label{eq:mbrunning}
\eeq
where $a(\mu)=\alpha_s(\mu)/\pi$,
$K = 1 + a(m_Z) \beta_0 \log(\mu^2/m_Z^2)$
with $\beta_0=(11-2/3 N_F)/4$, $\gamma_0=1$ and $N_F=5$ the number of
active flavours, to connect the running parameters at different scales.

Conversely, for a given value of $R_3^{b\ql}$
we can solve \eq{eq:pole} (or \eq{eq:MS}) with respect
to the quark mass. The result, shown in fig.~\ref{fig:radmu_dur}b
for $R_3^{b\ql}(\yc=0.02)=0.973$, depends on which equation was used and has
a residual dependence on the renormalization scale $\mu$.
The curves in fig.~\ref{fig:radmu_dur}b are obtained in the following way:
first from \eq{eq:MS} we directly obtain for an arbitrary value
of $\mu$ between $m_Z$ and $m_Z/10$ a value for the bottom-quark running
mass at that scale, $m_b(\mu)$, and then using \eq{eq:mbrunning}
we get a value for it at the $Z$-scale, $m_b(m_Z)$.
Second, using \eq{eq:pole} we extract, also for an arbitrary value
of $\mu$ between $m_Z$ and $m_Z/10$, a value for the
pole mass, $M_b$. Then we use \eq{eq:poltorunning} at $\mu=M_b$
and again \eq{eq:mbrunning} to perform the evolution from $\mu=M_b$
to $\mu=m_Z$ and finally get a value for $m_b(m_Z)$.
The two procedures give a different
answer since different higher orders have been neglected in the intermediate
steeps. The maximum spread of the two results,
in this case of the order of $\pm 200$~MeV, can be interpreted
as an estimate of the size of higher order corrections, that is,
of the theoretical error in the extraction of $m_b(m_Z)$
from the experimental measurement of $R_3^{b\ql}$.

\begin{figure}[t]
\begin{center}
\begin{tabular}{cc}
\psfig{figure=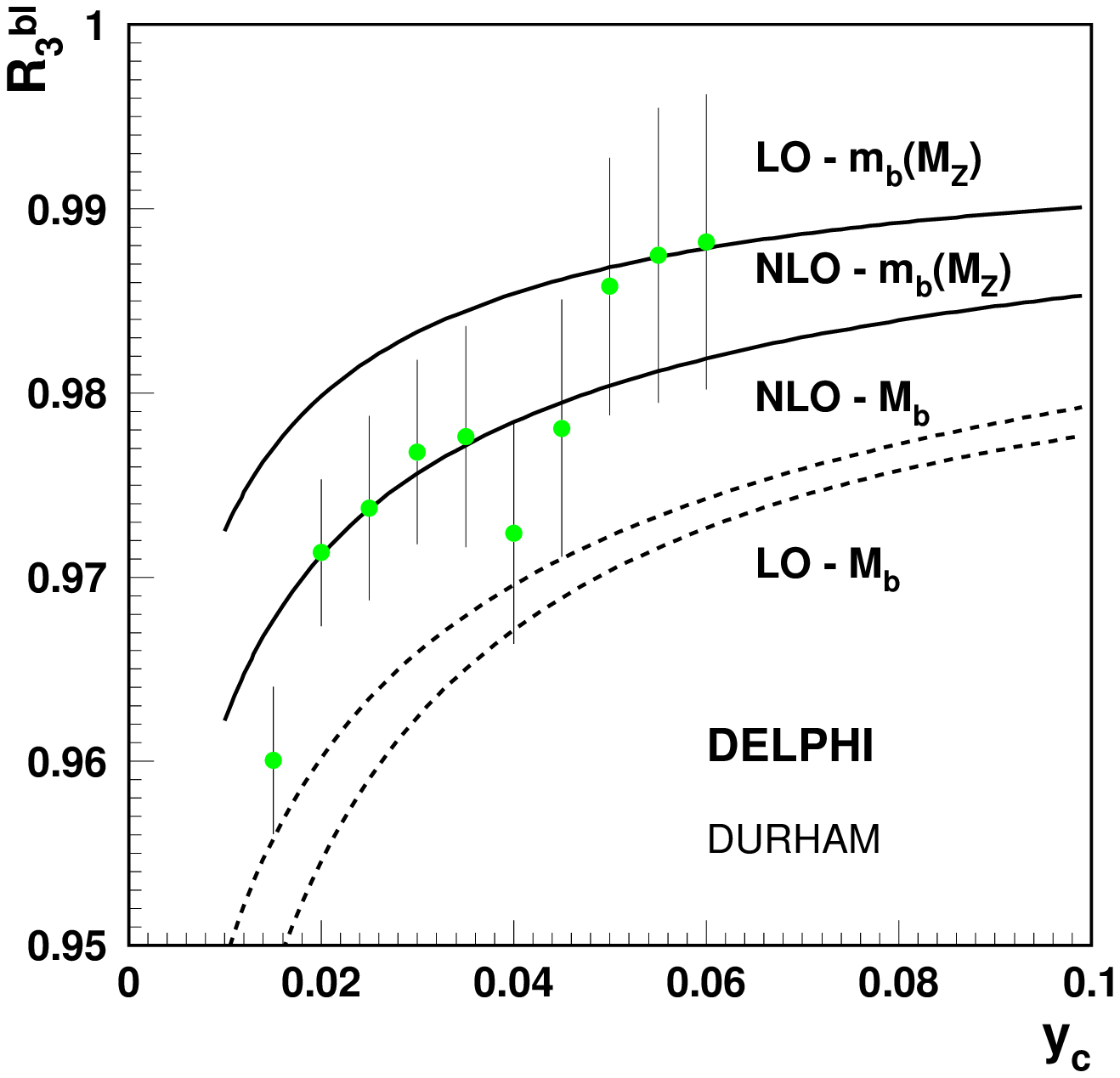,width=6. cm} &
\psfig{figure=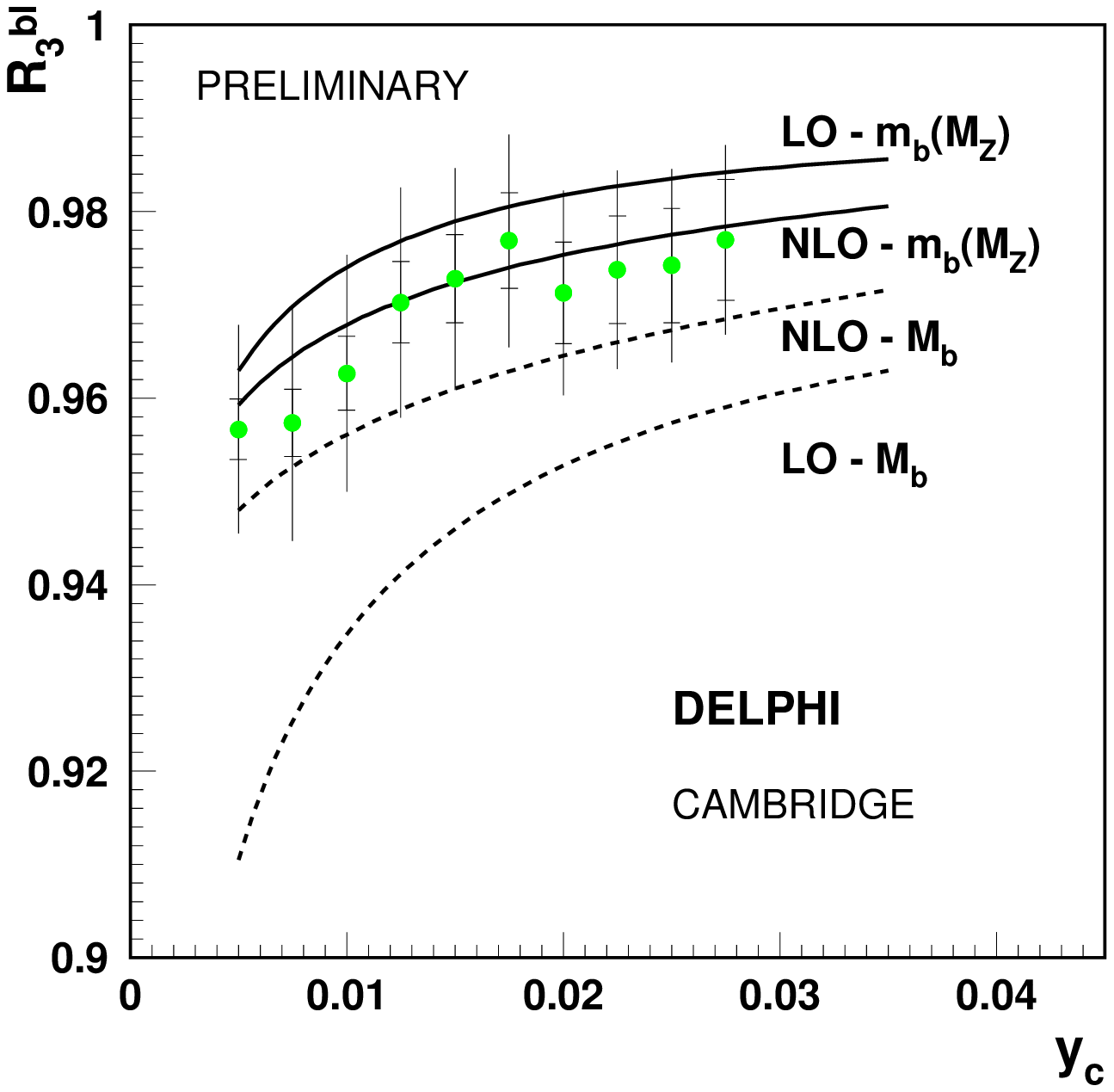,width=6. cm}
\end{tabular}
\end{center}
\caption{Experimental measurement of $R_3^{b\ql}$
by the DELPHI Coll. in the Durham algorithm and 
preliminary results in the Cambridge algorithm 
compared with our NLO calculation written
in terms of the running mass at the $Z$ peak, NLO-$m_b(m_Z)$,
or in terms of the pole mass, NLO-$M_b$. The LO predictions
are also plotted.
\label{fig:delphi_dur}}
\end{figure} 

These calculations were used by the DELPHI Coll.~\cite{delphi97}
to extract $m_b(m_Z)$ from the experimental measurement of
$R_3^{b\ql}$, see fig.~\ref{fig:delphi_dur}, and the
result was interpreted as the first experimental evidence
(at 2-3 sigmas) for the {\it running} of a fermion mass
since the data are better described by the NLO-$m_b(m_Z)$ curve.
Also recently the SLD Coll.~\cite{sld} has presented results for $m_b(m_Z)$.
The SLD analysis is compatible with the DELPHI measurement.
Nevertheless, the central values of $m_b(m_Z)$ obtained 
from different clustering algorithms are 
scattered in the range $\Delta m_b(m_Z) = \pm 0.49$~GeV.
This is probably due to the fact that E-like algorithms,
that are mainly used in this analysis, have huge NLO corrections
thus making accurate QCD predictions difficult.


\section{Tests of the flavour independence of the strong interaction}

Assuming a given $b$-quark mass the NLO calculation of heavy quark  
three-jet production cross section can be used to perform an
improved test of the flavour independence of the strong 
coupling constant. Such analysis was done for the first time by the 
DELPHI Coll. in~\cite{delphi97} by using the 
$R_3^{b\ql}$ observable defined in the Durham algorithm.
Recently, the OPAL Coll.~\cite{opal} has presented a similar analysis
by using different ratios of event shapes distributions: 
$D_2$, $1-T$, $M_H$, $B_T$, $B_W$ and $C$.
Instead, SLD has presented~\cite{sld}
results by analyzing the $R_3^{bl}$ ratio
in the E, E0, P, P0, Durham and Geneva algorithms.
All these results are consistent with unity. 
{\it No flavour dependence} has been observed.
Furthermore, the inclusion of mass effects is mandatory
to achieve such agreement.


\section{Improving the $b$-quark mass measurements:
the Cambridge algorithm}

The Cambridge algorithm has been introduced very
recently~\cite{Dokshitser:1997in} in order to reduce the
formation of spurious jets formed with low transverse momentum
particles that appear in the Durham algorithm at low $y_c$.
Therefore, compared to Durham, it allows to test smaller values
of $y_c$ while still keeping NLO corrections relatively small.
Both algorithms are defined by the same
recombination procedure and the same test variable
\beq
y_{ij} = 2 min(E_i^2,E_j^2)(1-\cos \theta_{ij})/s~,
\eeq
where $E_i$ and $E_j$ denote the energies of particles $i$ and $j$
and $\theta_{ij}$ is the angle between their momenta. 
The new ingredient of the Cambridge algorithm is the 
so called ordering variable 
\beq
v_{ij} = 2 (1-\cos \theta_{ij})~.
\eeq
In the Cambridge algorithm one first finds the minimal $v_{ij}$ and
then tests $y_{ij}$.
If $y_{ij}<y_c$, the $i$ and $j$ particles are recombined 
into a new pseudoparticle of momentum $p_k=p_i+p_j$
but if $y_{ij}>y_c$, the softer particle is resolved as a jet.
The net effect of the new definition is that
NLO corrections to the three-jet fraction become smaller.

In fig.~\ref{fig:delphi_dur} we present the preliminary
results from the DELPHI Coll.~\cite{juan98} for the
$R_3^{b\ql}$ ratio defined in the Cambridge algorithm and compare it 
with our NLO calculation~\cite{Rodrigo:1998nk} written in terms of the
running mass at the $Z$ peak, NLO-$m_b(m_Z)$, or in terms of the pole mass,
NLO-$M_b$, for $\mu=m_Z$ and $\alpha_s(m_Z)=0.118$.
As in Durham, the NLO-$m_b(m_Z)$ gives the best agreement. 
Furthermore, data are still compatible with the LO-$m_b(m_Z)$
showing that the bulk of higher order corrections is
described by the {\it running of the $b$-quark mass}.
In contrast, although data could also be well described by the
NLO-$M_b$ curve, the NLO corrections become large when 
the pole mass parametrization is used.

The studies of the NLO-$m_b(m_Z)$ curve show~\cite{Rodrigo:1998nk}
that it is remarkably stable with respect to the variation of the scale
$\mu$. For the range $m_Z/10 < \mu < m_Z$
the estimate of the error in the extracted $m_b(m_Z)$
is reduced to $\pm 50$~MeV in the Cambridge scheme,
with respect to $\pm 200$~MeV for Durham ($\pm 125$~MeV
if only NLO-$m_b(m_Z)$ is considered). In contrast,
when the NLO-$M_b$ parametrization is used we get $\pm 240$~MeV
but strongly dependent on the lower $\mu$ used.

\section{Conclusions} 

In the last few years an important progress was done in the description
of the $Z$-boson decay into three-jets with massive quarks.
Next-to-leading order calculations have been done by three groups
and have been successfully used in the analysis of the LEP and SLC
data where mass effects have been clearly seen.
Further studies of different observables and different jet-algorithms 
are oriented on the reduction of the theoretical uncertainty.
One good candidate might be the Cambridge jet-algorithm, 
where the NLO corrections are particularly small and where
the predictions in terms of the running mass, $m_b(m_Z)$
are particularly stable with respect to the variation of
the renormalization scale.

\section*{Acknowledgments}
We are very pleased to thank S. Cabrera, 
J. Fuster and S. Mart\'{\i} for an enjoyable collaboration. 
G.R. acknowledges a postdoctoral fellowship from INFN (Italy).
Work supported in part by CICYT (Spain), AEN-96-1718 and
DGESIC (Spain), PB97-1261.

\end{document}